\documentclass[12pt]{article}
\begin{document}
\begin{center}
\large\bf{Quantization of the First-Order \\
Two-Dimensional Einstein-Hilbert Action}
\end{center}
\vspace{2cm}
\begin{center}
D.G.C. McKeon\\
Department of Applied Mathematics\\
University of Western Ontario\\
London\\
CANADA\\
N6A 5B7
\end{center}
\vspace{1cm}
email:  dgmckeo2@uwo.ca \hfill PACS No. 11-15q\\
Tel: (519)661-2111, ext. 88789\\
Fax: (519)661-3523\vspace{3cm}\\

\section{Abstract}
A canonical analysis of the first-order two-dimensional Einstein-Hilbert action has shown it to have no physical degrees of freedom and to possess an unusual gauge symmetry with a symmetric field $\xi_{\mu\nu}$ acting as a gauge function. Some consequences of this symmetry are explored. The action is quantized and it is shown that all loop diagrams beyond one-loop order vanish. Furthermore, explicit calculation of the one-loop two-point function shows that it too vanishes, with the contribution of the ghost loop cancelling that of the ``graviton'' loop.
\eject

The two-dimensional Einstein-Hilbert action
$$S = \int d^2x \sqrt{-g} \,g^{\mu\nu} R_{\mu\nu}\eqno(1)$$
is a purely topological action provided the Ricci curvature tensor $R_{\mu\nu}$ is written as
$$R_{\mu\nu} = \left\lbrace \begin{array}{c} \lambda \\ \mu\nu \end{array}\right\rbrace_{,\lambda} - 
\left\lbrace \begin{array}{c} \lambda \\ \lambda\mu \end{array}\right\rbrace_{,\nu}
+ \left\lbrace \begin{array}{c} \lambda \\ \mu\nu \end{array}\right\rbrace
\left\lbrace \begin{array}{c} \sigma \\ \sigma\lambda \end{array}\right\rbrace
- \left\lbrace \begin{array}{c} \lambda \\ \sigma\mu \end{array}\right\rbrace
\left\lbrace \begin{array}{c} \sigma \\ \lambda\nu \end{array}\right\rbrace \eqno(2)$$
with the Christoffel symbol $\left\lbrace \begin{array}{c} \lambda \\ \mu\nu \end{array}\right\rbrace$ being given by 
$$\left\lbrace \begin{array}{c} \lambda \\ \mu\nu \end{array}\right\rbrace = \frac{1}{2} g^{\lambda \rho} \left( g_{\rho\mu , \nu} + g_{\rho \nu , \mu} - g_{\mu\nu , \rho}\right)\eqno(3)$$
when expressed in terms of the metric $g_{\mu\nu}$. (The flat space metric $\eta^{\mu\nu}$ is taken to be $\eta^{00} = -\eta^{11} = -1$; we also use the antisymmetric matrix $\epsilon^{\mu\nu}$ with $\epsilon^{01} = -\epsilon^{10} = 1$.)

In refs. [1-6] a first-order form of eq. (1) is considered. This entails expressing $R_{\mu\nu}$ not as a function of the metric through eqs. (2) and (3), but rather as a function of an independent affine connection $\Gamma_{\mu\nu}^\lambda = \Gamma_{\nu\mu}^\lambda$ with $\Gamma_{\mu\nu}^\lambda$ replacing $\left\lbrace \begin{array}{c} \lambda \\ \mu\nu \end{array}\right\rbrace$ in eq. (2). If we define
$$h^{\mu\nu} = \sqrt{-g}\,g^{\mu\nu} \eqno(4)$$
and
$$G_{\mu\nu}^\lambda = \Gamma_{\mu\nu}^\lambda - \frac{1}{2} \left( \delta_\mu^\lambda \Gamma_{\sigma\nu}^\sigma + \delta_\nu^\lambda \Gamma_{\sigma\mu}^\sigma\right)\eqno(5)$$
then the action of eq. (1) when generalized to $d$ dimensions becomes
$$S = \int d^dx h^{\mu\nu} \left(G_{\mu\nu ,\lambda}^\lambda + \frac{1}{d - 1} G_{\lambda\mu}^\lambda G_{\sigma\nu}^\sigma - G_{\sigma\mu}^\lambda G_{\lambda\nu}^\sigma\right).\eqno(6)$$
Treating $h^{\mu\nu}$ and $G_{\mu\nu}^\lambda$ as independent fields, a canonical analysis of eq. (6) shows that when $d = 2$ it is invariant under the infinitesimal transformation
$$\delta h^{\mu\nu} = \left(\epsilon^{\mu\lambda} h^{\sigma\nu} + \epsilon^{\nu\lambda} h^{\sigma\mu}\right)\xi_{\lambda\sigma}\eqno(7)$$
$$\delta G_{\mu\nu}^\rho = - \epsilon^{\rho\lambda} \xi_{\mu\nu , \lambda} - \epsilon^{\lambda\sigma}\left(G_{\mu\lambda}^\rho \xi_{\sigma\nu} + G_{\nu\lambda}^\rho \xi_{\sigma\mu}\right)\eqno(8)$$
where $\xi_{\mu\nu} = \xi_{\nu\mu}$ [5].  Taking $h_{\mu\nu}$ to be the inverse of $h^{\mu\nu}$ so that $h_{\mu\lambda} h^{\lambda\nu} = \delta_\mu^\nu$, then eq. (7) implies that
$$\delta h_{\mu\nu} = -\epsilon^{\lambda\sigma}\left(h_{\lambda\mu} \xi_{\sigma\nu} + h_{\lambda\nu}\xi_{\sigma\mu}\right).\eqno(9)$$
From eq. (6) when $d = 2$, the equations of motion for $h_{\mu\nu}$ and $G_{\mu\nu}^\rho$ are respectively
$$R_{\mu\nu} = G_{\mu\nu , \lambda}^\lambda + G_{\lambda\mu}^\lambda G_{\sigma\nu}^\sigma - G_{\sigma \mu}^\lambda G_{\lambda\nu}^\sigma = 0\eqno(10)$$
and
$$- h^{\mu\nu}_{\;\;, \rho} + G_{\lambda\sigma}^\lambda \left(\delta_\rho^\mu h^{\sigma\nu} + \delta_\rho^\nu h^{\sigma\mu}\right)
- \left(G_{\rho\sigma}^\mu h^{\sigma\nu} + G_{\rho\sigma}^\nu h^{\sigma\mu}\right) = 0.\eqno(11)$$
Unlike the situation when $d \neq 2$, we cannot solve for $G_{\;\;\rho}^{\mu\nu}$ in terms of $h^{\mu\nu}$ using eq. (11), as multiplying eq. (10) by $h_{\mu\nu}$ does not result in an equation for $G_{\lambda\sigma}^\lambda$, but rather the identity
$h_{\mu\nu} h^{\mu\nu}_{\;\;, \lambda} = 0$. Eq. (11) is satisfied by
$$G_{\mu\nu}^\lambda = \frac{1}{2} h^{\lambda\rho}\left(h_{\rho\mu , \nu} + h_{\rho\nu ,\mu} - h_{\mu\nu , \rho}\right) + h_{\mu\nu} X^{\lambda},\eqno(12)$$
where $X^\lambda$ is an arbitrary vector [1-4]. The action of eq. (6) when $d = 2$ does not receive any contribution from $X^\lambda$. From eq. (12) it follows that
$$X^\mu = h^{\mu\nu} G_{\lambda\nu}^\lambda.\eqno(13)$$
Under the transformations of eqs. (7) and (8), the change in $X^\mu$ is thus given by
$$\delta X^\mu = \delta h^{\mu\nu} G_{\lambda\nu}^\lambda + h^{\mu\nu}\delta G_{\lambda\nu}^\lambda ;\eqno(14)$$
eqs. (7), (8) and (14) are consistent with eq. (12).

The transformations of eqs. (7) and (8) are distinct from a diffeomorphism transformation in two dimensions.  This is immediately apparent, as the gauge parameter $\xi_{\mu\nu}$ has three independent components, while a diffeomorphism is characterized by just two parameters.  Coupling either $h^{\mu\nu}$ or $G_{\mu\nu}^\lambda$ to a matter field in a way that respects the symmetry associated with eqs. (7) and (8) does not appear to be feasable.

From eq. (4), we see that in $d$ dimensions
$$h \equiv \det h^{\mu\nu} = -(-g)^{\frac{d-2}{2}}.\eqno(15)$$
If $d = 2$, eq. (13) becomes a constraint equation
$$h = -1.\eqno(16)$$
This constraint is respected by the transformation of eq. (7). When $d = 2$, it is not possible to express $g^{\mu\nu}$ in terms of $h^{\mu\nu}$ on account of eq. (16), and consequently we cannot supplement the action of eq. (6) with a cosmological term $\Lambda \sqrt{-g}$ when $d = 2$ if we wish to work in terms of the fields $h^{\mu\nu}$. However, if we ensure that the constraint of eq. (16) is satisfied by adding [5]
$$S_\lambda = \int d^dx \lambda (h + \kappa)\eqno(17)$$
to the action of eq. (6) when $d = 2$ ($\lambda$ being a Lagrange multiplier and $\kappa$ being a field that can be set equal to one), the cosmological term can be taken to be $\Lambda (\kappa)^{\frac{1}{d-2}}$ in the limit $d \rightarrow 2$.

In order to quantize the action of eq. (6) when $d = 2$, we choose the gauge fixing action
$$S_{gf} = \int d^2x \left(\frac{-1}{2\alpha}\right)\left(\epsilon_{\mu\nu} G_{\alpha\beta}^{\mu ,\nu}\right) \left(\epsilon^{\lambda\sigma} G_{\lambda , \sigma}^{\alpha\beta}\right).\eqno(18)$$
One could retain the first order form of the action by introducing a Nakanishi-Lautrup field $N^{\alpha\beta}$ and using the gauge fixing action
$$S_{gf} = \int d^2x \left(-N^{\alpha\beta} \epsilon_{\lambda\sigma} G_{\alpha\beta}^{\lambda ,\sigma} + \frac{\alpha}{2} N_{\alpha\beta} N^{\alpha\beta}\right).\eqno(19)$$
For the ensuing discussion, we will use eq. (18) for $S_{gf}$; the same results we will present for the two-point function follow from eq. (19).

The bilinear terms in the Lagrangian formed by adding eqs. (6) and (18) together are of the form
$$\textit{L}^{(2)} = \frac{1}{2}\left(h^{\mu\nu}, G_\lambda^{\alpha\beta}\right)
\left(
\begin{array}{cc}
0 & \Delta_{\mu\nu}^{\gamma\delta} \partial_\sigma\\
-\Delta_{\alpha\beta}^{\pi\tau} \partial^\lambda &
-\frac{1}{\alpha} \partial^2 T_\sigma^\lambda\Delta_{\alpha\beta}^{\gamma\delta}\end{array}\right)\left(\begin{array}{c}
h_{\pi\tau}\\
G_{\gamma\delta}^\sigma\end{array}\right).\eqno(20)$$
The inverse of the matrix $\begin{array}{c}M\\[-4.90mm]
\!\!\sim
\end{array}$ appearing in eq. (20) is
$$\begin{array}{c}
M^{-1}\\[-4.90mm]
\!\!\!\!\!\!\sim
\end{array}
 = \frac{1}{\partial^2}
\left(\begin{array}{cc}
0 & -\Delta_{\pi\tau}^{\xi\zeta} \partial_\kappa\\
+\Delta_{\gamma\delta}^{\epsilon\rho} \partial^\sigma &
-\alpha T_{\phi}^\sigma\Delta_{\gamma\delta}^{\xi\zeta}\end{array}\right).\eqno(21)$$
(We use the notation $
\Delta_{\mu\nu}^{\alpha\beta} = \frac{1}{2}\left(\delta_\mu^\alpha \delta_\nu^\beta + \delta_\nu^\alpha\delta_\mu^\beta\right)$ and
$T_\mu^\alpha = \delta_\mu^\alpha - \partial^\alpha\partial_\mu/\partial^2$.)
From eq. (19) we obtain the propagators
$$\left\langle G_{\alpha\beta}^\sigma (-k) G_\kappa^{\gamma\delta}(k)\right\rangle = \frac{i\alpha}{k^2} T_\kappa^\sigma \Delta_{\alpha\beta}^{\gamma\delta}\eqno(22)$$
$$\left\langle G_{\alpha\beta}^\sigma (-k) h^{\gamma\delta}(k)\right\rangle = \frac{k^\sigma}{k^2} \Delta_{\alpha\beta}^{\gamma\delta}\eqno(23)$$
$$\left\langle h_{\alpha\beta}(-k) G_\kappa^{\gamma\delta}(k)\right\rangle = -\frac{k_\kappa}{k^2} \Delta_{\alpha\beta}^{\gamma\delta}\eqno(24)$$
$$\left\langle h_{\alpha\beta}(-k) h^{\gamma\delta}(k)\right\rangle = 0 \,.\eqno(25)$$
In eqs. (20-22), $k^\mu$ is the momentum associated with the propagator, taken as flowing towards the field appearing to the right within the bracket $\left\langle\;\; \right\rangle$. Furthermore, from the action of eq. (6), we obtain the vertex 
$$\begin{array}{lrr}
\;\;\;\;\;\;\left\langle h^{\mu\nu} G_{\alpha \beta}^\lambda G_{\gamma\delta}^\sigma\right\rangle &
= \frac{i}{2}
\left[ - \delta_\gamma^\lambda \delta_\alpha^\sigma \Delta^{\mu\nu}_{\beta\delta}\right. &
- \delta_\gamma^\lambda \delta_\beta^\sigma \Delta^{\mu\nu}_{\alpha\delta} \\
& - \delta_\delta^\lambda \delta_\alpha^\sigma \Delta^{\mu\nu}_{\beta\gamma} & -\delta_\delta^\lambda \delta_\beta^\sigma \Delta^{\mu\nu}_{\alpha\gamma}\\
& + \delta_\alpha^\lambda \delta_\gamma^\sigma \Delta^{\mu\nu}_{\beta\delta} &
+ \delta_\beta^\lambda \delta_\gamma^\sigma \Delta^{\mu\nu}_{\alpha\delta}\\
& + \delta_\alpha^\lambda \delta_\delta^\sigma \Delta^{\mu\nu}_{\beta\gamma} &
\left.+ \delta_\beta^\lambda \delta_\delta^\sigma \Delta^{\mu\nu}_{\alpha\gamma}\right]\end{array}\eqno(26)$$
$$\equiv X_{\alpha\beta , \gamma\delta}^{\mu\nu ,\lambda , \sigma}.\nonumber$$

The gauge transformation of eqs. (7) and (8) imply the necessity of including contributions of Faddeev-Popov ghost fields. The gauge condition associated with eq. (16) is
$$\epsilon_{\mu\nu} G_{\alpha\beta}^{\mu ,\nu} = 0,\eqno(27)$$
which, when subject to the variation of eq. (7), leads to the Faddeev-Popov ghost Lagrangian
$$\textit{L}_{gh} = -\overline{\zeta}^{\alpha\beta}(\epsilon_{\mu\nu})\left[ -\epsilon^{\mu\rho} \zeta_{\alpha\beta ,\rho} - \epsilon^{\rho\sigma} \left(G_{\alpha \rho}^\mu \zeta_{\sigma\beta} + G_{\beta \rho}^\mu \zeta_{\sigma\alpha}\right)\right]^{,\nu}\eqno(28)$$
where $\zeta_{\alpha\beta}$ and $\overline{\zeta}_{\alpha\beta}$ are the ghost fields. Using the identity $\epsilon^{\lambda\sigma} \epsilon_{\mu\nu} = - \delta_\mu^\lambda \delta_\nu^\sigma + \delta_\nu^\lambda \delta_\mu^\sigma$, eq. (26) reduces to 
$$\textit{L}_{gh} = -\overline{\zeta}^{\alpha\beta}\partial^2\zeta_{\alpha\beta} + 2 \overline{\zeta}^{\alpha\beta ,\sigma}\left( G_{\alpha\rho}^\rho \zeta_{\sigma\beta} - G_{\alpha\sigma}^\rho \zeta_{\rho\beta}\right).\eqno(29)$$
Consequently, the ghost propagator is
$$\left\langle \zeta_{\alpha\beta}(-k)\overline{\zeta}^{\gamma\delta}(-k)\right\rangle
= \frac{i}{k^2}\Delta_{\alpha\beta}^{\gamma\delta}\eqno(30)$$
and the vertex involving ghosts is
$$\left\langle G_{\mu\nu}^\lambda (-p-q)\overline{\zeta}^{\alpha\beta}(p)\zeta_{\gamma\delta} (q)\right\rangle
= -\frac{1}{2}
\left[ \Lambda_{\gamma\nu}^\lambda (p) \Delta_{\mu\delta}^{\alpha\beta} + \Lambda_{\gamma\mu}^\lambda (p) \Delta_{\nu\gamma}^{\alpha\beta}\right.\nonumber$$
$$\left. +\Lambda_{\delta\nu}^\lambda (p) \Delta_{\mu\gamma}^{\alpha\beta} + \Lambda_{\delta\mu}^\lambda (p) \Delta_{\nu\delta}^{\alpha\beta}\right]\eqno(31)$$
$$\equiv Y_{\mu\nu ,\gamma\delta}^{\lambda , \alpha\beta}(p)\nonumber$$
where $\Lambda_{\gamma\nu}^\lambda (p) \equiv (p_\gamma \delta_\nu^\lambda - p_\nu\delta_\gamma^\lambda)$ with $p$ being the momentum flowing into the field $\overline{\zeta}^{\alpha\beta}$ away from the vertex.

In the ``Landau gauge'' in which $\alpha = 0$, the only non-vanishing propagators are the mixed propagators of eqs. (23) and (24). This, combined with the fact that the only vertices are $\left\langle hGG \right\rangle$ and 
$\left\langle G\overline{\zeta}\zeta \right\rangle$, make it easy to establish that the only possible Feynman diagrams are restricted to being one-loop, much as in the model of ref. [7].

If now with $\alpha = 0$, we compute the one-loop diagrams in which there are external fields $G_{\alpha\beta}^\lambda(-p)$ and $G_{\gamma\delta}^\sigma (p)$, the mixed propagators of eqs. (21) and (22) contribute to the ``graviton"' loop
$$I_{1\;\lambda , \sigma}^{\alpha\beta , \gamma\delta} (p) = \int \frac{d^2k}{(2\pi)^2} \left(\frac{(k+p)^\xi}{(k+p)^2} \Delta_{\epsilon\kappa}^{\mu\nu}\right)\left(\frac{k^\zeta}{k^2} \Delta_{\omega\phi}^{\pi\tau}\right)\eqno(32)$$
$$\left(X_{\pi\tau ,\lambda , \xi}^{\alpha\beta ,\epsilon\kappa}\right)\left(X_{\mu\nu ,\sigma ,\zeta}^{\gamma\delta , \omega\phi}\right),\nonumber$$
while the ghost loop involving the propagator of eq. (28) contributes
$$I_{2\;\lambda , \sigma}^{\alpha\beta , \gamma\delta}= - \int \frac{d^2k}{(2\pi)^2} \left(\frac{i}{k^2} \Delta_{\omega\phi}^{\pi\tau}\right)\left(\frac{i}{(k+p)^2} \Delta_{\epsilon\kappa}^{\mu\nu}\right)
Y_{\lambda ,\epsilon\kappa}^{\alpha\beta ,\pi\tau} (k+p) 
Y_{\sigma , \mu\nu}^{\gamma\delta , \omega\phi}(k).\eqno(33)$$
Since $\Delta_{\beta\sigma}^{\alpha\lambda} \Delta_{\delta\lambda}^{\gamma\sigma} = \delta_\beta^\alpha \delta_\delta^\gamma + \frac{1}{4} \delta_\beta^\gamma \delta_\delta^\alpha$, eq. (30) reduces to
$$I_{1\;\lambda , \sigma}^{\alpha\beta , \gamma\delta} =  \frac{1}{2} \int
\frac{d^2k}{(2\pi)^2}
\frac{1}{(k+p)^2k^2} \left[
(k+p)^\alpha k^\gamma  \Delta_{\lambda\sigma}^{\beta\delta}\right.\nonumber$$
$$+ (k + p)^\alpha k^\delta \Delta_{\lambda\sigma}^{\beta\gamma} + (k + p)^\beta k^\gamma \Delta_{\lambda\sigma}^{\alpha\delta}\nonumber$$
$$\left. + (k + p)^\beta k^\delta \Delta_{\lambda\sigma}^{\alpha\gamma} -2 (k + p)^\alpha k^\beta \Delta_{\lambda\sigma}^{\delta\gamma} - 2 (k +p)^\gamma k^\delta \Delta_{\lambda\sigma}^{\alpha\beta}\right]\eqno(34)$$
while from eq. (33), $I_{2\lambda , \sigma}^{\alpha\beta ,\gamma\delta} = -I_{1 , \lambda ,\sigma}^{\alpha\beta ,\gamma\delta}$. This ensures that there is no net radiative correction to this two point function. The cancellation between $I_1$ and $I_2$ occurs without having to choose a regulating procedure to define divergent integrals.

Having no radiative corrections to the two point function is distinct from what occurs when considering the action of eq. (1) with $R_{\mu\nu}$ given by eqs. (2) and (3). In this case, since the classical action is a pure surface term, the radiative corrections receive contributions solely from the ghost loop, with the ghost fields being associated with the differeomorphism and conformal invariance of eqs. (1-3), rather than the gauge symmetry of eqs. (7-8) [8,9]. This ghost loop results in  a radiative correction to the propagor for the metric being given by [10-12] the bilinear contribution to the induced gravitational action
$$\Gamma (g) =\int d^2x \left(\frac{13}{24} R\, \partial^{-2} \,R\right).\eqno(35)$$
Consequently, radiative corrections to the first and second order form of the two dimensional Einstein-Hilbert action are different.

\section{Acknowledgements}

We would like to thank N.Kiriushcheva and S. Kuzmin for discussions.  NSERC provied financial support.  Roger Macloud had a helpful suggestion.

\end{document}